\documentclass{iau}
\usepackage{graphicx}
\usepackage{amsmath}
\usepackage{multirow}
\usepackage{hyperref}
\usepackage{natbib}

\title[Intraday variations of polarization vector in blazars] 
{Intraday variations of polarization vector in blazars: a key to the optical jet structure?}

\author[Shablovinskaya et al.]   
{Elena Shablovinskaya, Eugene Malygin, Dmitry Oparin}

\affiliation{Special Astrophysical Observatory of RAS (SAO RAS), 369167 {Nizhny Arkhyz}, Russia}

\pubyear{2023}
\volume{375}  
\jname{The multimessenger chakra of blazar jets}
\editors{I. Liodakis, M.F. Aller, \\ H. Krawczynski, A. Lähteenmäki, \& T.J. Pearson, eds.}
\begin{document}

\maketitle

\begin{abstract}
This report presents the results of optical polarimetric observations carried out with 6-m and 1-m telescopes at SAO RAS. The study of the blazar S5 0716+714 radiation showed the presence of a period of the variability of brightness and polarization vector variations on scales of $\sim$1.5 hours, constant on a long time scale; multi-colour monitoring of BL Lac polarization before, during and after the flare demonstrates the difference in the patterns of polarization vector variability depending on the wavelength. Several geometrical models and physical descriptions are discussed.
\end{abstract}

\firstsection 
\section{Introduction}

For several decades, the optical variability of blazars as a special type of active galactic nuclei (AGNs) with the jet oriented towards the observer has been actively investigated. The most popular is the accumulation of long-term photometric data series \cite[e.g.][]{robopol}. However, the variability of blazars is violent and stochastic; and just as it is impossible to unambiguously reconstruct the three-dimensional air masses flow in the Earth's atmosphere according to the graph of wind speed changes over time, it is difficult to understand exclusively by the blazar light curves what physical and dynamical processes the plasma is driven by.

Fortunately, blazars are highly polarized in all spectral ranges. Their polarization of a synchrotron nature is related both to the dynamics of the emitting plasma in the unresolved jet region at scales $<$0.01 pc from the nucleus and to the magnetic field in which the plasma moves. This initiates curiosity to study and map the rotation of the polarization vector, which is expected to be fast due to relativistic velocities and the small linear size of the jet.

Intraday variations (IDV) of the polarization vector direction are observed in the optical and radio bands. In the case of radio observations, IDV clearly showed the helical structure of the jet \citep[e.g.][]{cta102}. Patterns of optical polarization are not so clear to unambiguously reveal the same; however, long-term intraday polarization monitoring in perspective is a good tool for checking the inner models of the optical jet. Moreover, IDV continues to be little studied to date, which leaves a number of questions, e.g., how IDV depends on activity states and what physical processes dominate. 
Raising such issues, for several years we have been conducting intraday monitoring of the polarization of a sample of blazars. To date, we have obtained the most interesting results for two of them: S5 0716+714 and BL Lac. We briefly describe these results below.

\section{Brief comments on observational technique}

The observations were conducted at the 6-m BTA with SCORPIO-2 \citep{scorpio2} and 1-m Zeiss-1000 of SAO RAS with StoP \citep{stop} and MAGIC \citep{magic_an} devices. Regardless of the device, for polarimetric observations, we use double Wollaston prism \citep{oliva,geyer} to conduct observations in the so-called one-shot polarimetry mode unaltered by rapid atmosphere flickering. Thus, the Stokes parameters $I$, $Q$, $U$ are measured independently for each frame. Combining it with the differential measurements by local standard stars in the field we achieve high accuracy of polarimetry (typically 0.1\% for AGNs). The detailed description of prisms parameters and observational and reduction techniques could be found in \cite{AfAm,0716,bllac}.

\section{S5 0716+714 and geometrical model}

\begin{figure}
    \centering
\includegraphics[scale=0.5,angle=90]{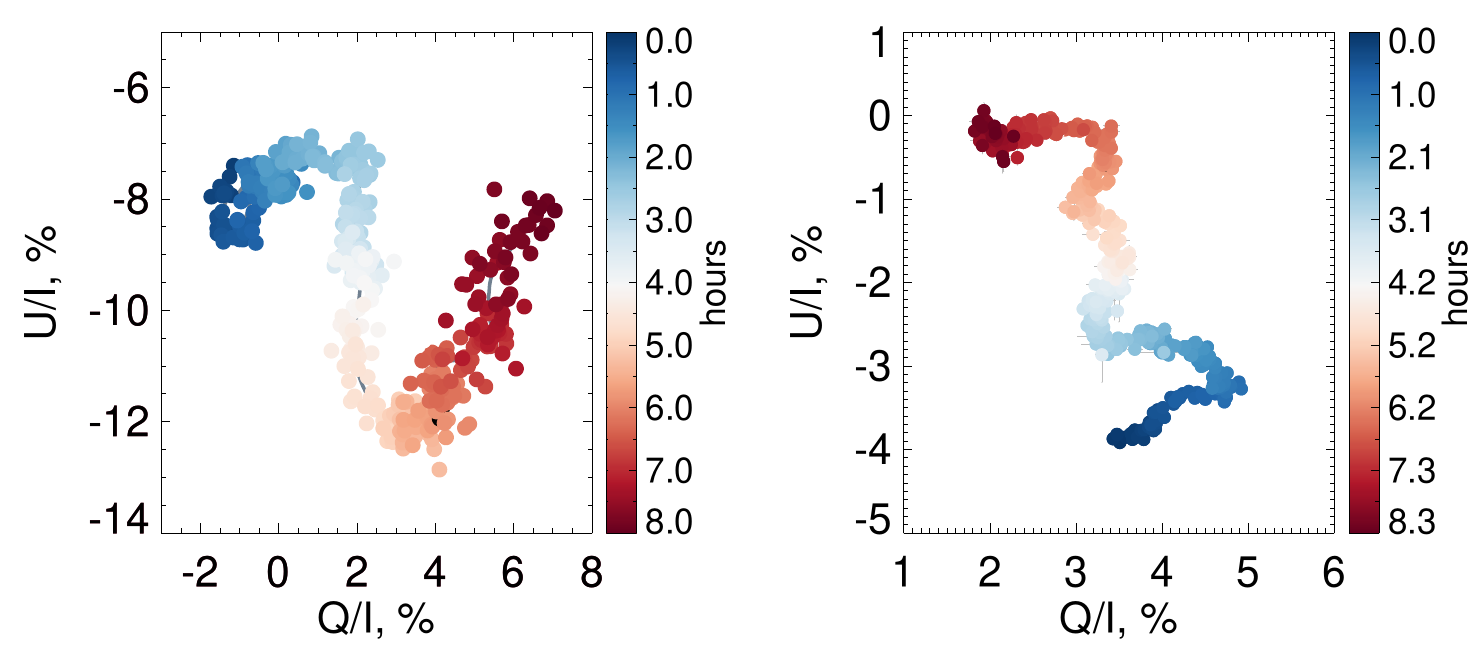}
    \caption{Overnight variations of the normalized Stokes parameters $Q$ and $U$ projected onto the $QU$-plane in S5 0716+714: 2018 \citep[left,][]{0716} and 2020 \citep[right,][]{stop} observation runs.}
    \label{0716}
\end{figure}

S5 0716+714 is one of the brightest and most variable blazars of the northern sky, attracting special attention due to the lack of unambiguous estimates of its redshift. For the first time, we conducted its polarimetric monitoring with a 1 min cadence during the whole night on the 6-m telescope, which allowed us to detect the polarization vector changes on small time scales: switches occurred with a period of $\sim$1.5 hours, and the same period was demonstrated by total flux variations \citep[more details in][]{0716}. This corresponds to the cross-section size of an optical jet of the order of 10 a.u. Two years later, we repeated similar observations of S5 0716+714 on the 1-m telescope (initially for testing a new device). The analysis of the latest data confirmed the result obtained earlier at the BTA and showed that the size of the jet emitting in the optical range is stable\footnote{or, at least, typical if we were able to detect it in observations spaced in time by several years.} \citep{stop}. Variations of the polarization vector on the $QU$-plane are given in Fig. \ref{0716}.

To explain the rotation of the polarization vector based on the works by \citet{steffen95,nalewajko09}, we have constructed a simple geometric model of plasma motion in a helical magnetic field in a conical jet. It is shown that the behaviour of the polarization vector on the $QU$-plane can be described by this model taking into account the precession of the magnetic field with a period of $\sim$15 days. Moreover, following the calculations of \citet{but20}, observed magnitude variations are predicted due to the Doppler factor changes. However, such a simple model is not so satisfactory because it resembles Ptolemy's epicycles: the more parameters are laid down, the more complex (and in general -- any) pattern of polarization variations can be obtained. This prompted us to expand our observations to be able to test physical processes in plasma.

\section{BL Lac and physical insights}

For S5 0716+714, the $QU$-plane revealed the complex and smooth polarization IDV trajectories. But what would the picture look like in different colours? Generally speaking, due to synchrotron losses, the patterns of polarization variability should be the same, but slightly shifted in time, and this shift is proportional to the magnetic field strength. In the case of the contribution of external polarization mechanisms, the trajectories will be stably shifted relative to each other; if the physical processes in the plasma are radically different in different optical bands, then the behaviour of the polarization vector in different colours will not correlate.

\begin{figure}
    \centering
    \includegraphics[scale=0.55,angle=90]{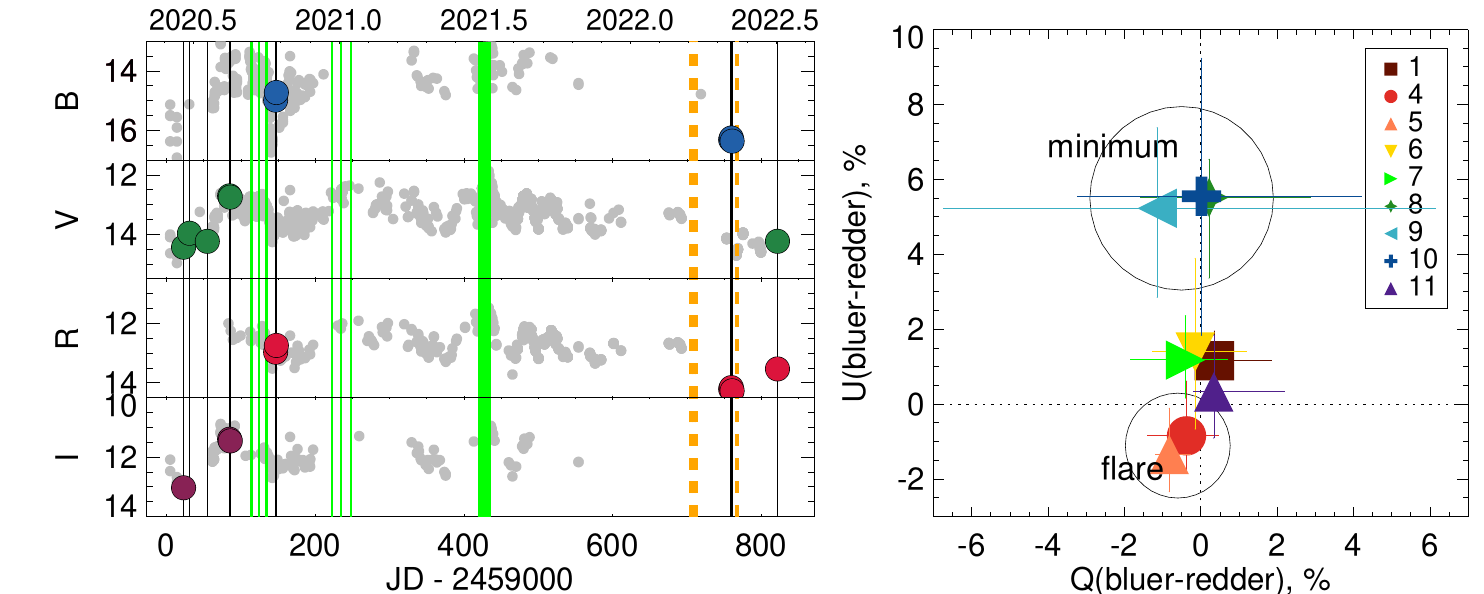}
    \caption{Left: BL Lac light curves according to \href{https://www.aavso.org /}{AAVSO} data. Our monitoring data is marked with circles. The green vertical bars indicate optical observations from \citep{imazawa22}, and orange dashed lines are for \textit{IXPE} epochs from \citep{middei22}. Right: $QU$-diagram for the polarization difference between the "redder"{} and "bluer"{} bands \citep[details in][]{bllac}}
    \label{bllac}
\end{figure}

To test this, we carried out observations in two optical filters (conditionally "red"{} and conditionally "blue"{}) of the blazar BL Lac in 2020-2022, when the object showed extraordinary variability. Due to the activity of the blazar, other authors also observed its multi-wavelength polarization during the same period, but our epochs did not coincide, although were often close (see Fig. \ref{bllac}, left).

In 4 of the 9 epochs, when the blazar was in a relatively quiet state, we observed variations of brightness and polarization vector rotations without a pronounced period. The polarization chromatism was moderate, with the dominance of the "blue"{} component. However, during the two extreme states of the blazar, the picture was different: during the flare, the polarization vector had sharper rotation but was weaker of $\sim$1-8\% and dominantly "red"{}. During the deep minimum of the blazar, on the contrary, the variability in integral and polarized light was weak, but the maximum "blue"{} chromatism with a degree of polarization up to $\sim$30\% was observed [see Fig. \ref{bllac}, on the right and \citep{bllac} for details].

The obtained observational data allow us to draw some conclusions (i.e. exclude some models) about the physical processes inside the optical jet. First, due to the polarization IDV and its chromatism on scales of an hour, assumptions about the external nature of the dependence of polarization on wavelength are excluded. In particular, any influence of the accretion disk is excluded due to the high polarization degree in the minimum state. To explain polarization chromatism, it is possible to assume processes in a synchrotron emitting plasma, but then the electron energy distribution must be broken and change at short times. Qualitatively, the polarization behaviour can be described by the model of turbulent plasma with a shock \citep{marscher21} or magnetic reconnection \citep{zhang22}. But in order to determine which acceleration process takes place, new numerical models of both processes are required.

\section{Further perspectives}

By S5 0716+714 and BL Lac observations, it can be seen the polarization IDV can not only resolve the plasma motion in the optical jet but also in the future be a critical test for physical models of plasma acceleration and emission. However, this still requires more extensive statistics, including different phases of activity of blazars of different types, at larger time scales and in a wider spectral range. On the other hand, the observations demonstrate the need to combine models of synchrotron radiation, the evolution of turbulent cells and the magnetic field, etc. 


{ \small Observations with the SAO RAS telescopes are supported by the Ministry of Science and Higher Education of the Russian Federation. The renovation of telescope equipment is currently provided within the national project "Science and Universities". We obtained part of the observed data on the unique scientific facility "Big Telescope Alt-azimuthal"{} of SAO RAS as well as made data processing with the financial support of grant No075-15-2022-262 (13.MNPMU.21.0003) of the Ministry of Science and Higher Education of the Russian Federation.}


\end{document}